\documentclass[preprint,aps,tightenlines,showpacs,nofootinbib]{revtex4}
\usepackage{latexsym}
\usepackage[dvips]{graphicx}
\newcommand{\beq}{\begin{eqnarray}}
\newcommand{\eeq}{\end{eqnarray}}

\newcommand{\ci}{\cite}

\newcommand{\om}{{\omega}}
\newcommand{\Om}{{\Omega}}

\newcommand{\wt}{\widetilde}
\begin{document}

\preprint{hep-th/0602114}

\title{Thermodynamics of Exotic Black Holes, Negative Temperature, and
Bekenstein-Hawking Entropy}

\author{Mu-In Park\footnote{E-mail address: muinpark@yahoo.com}}

\affiliation{ Center for Quantum Spacetime,  Sogang University,
Seoul 121-742, Korea}

\begin{abstract}
Recently, exotic black holes whose masses and angular momenta are
interchanged have been found, and it is known that their entropies
depend only on the $inner$ horizon areas. But a basic problem of
these entropies is that the second law of thermodynamics is not
guaranteed, in contrast to the Bekenstein-Hawking entropy. Here, I
find that there is another entropy formula which recovers the usual
Bekenstein-Hawking form, but the characteristic angular velocity and
temperature are identified with those of the inner horizon, in order
to satisfy the first law of black hole thermodynamics. The
temperature has a $negative$ value, due to an upper bound of mass as
in spin systems, and the angular velocity has a $lower$ bound. I
show that one can obtain the same entropy formula from a conformal
field theory computation, based on classical Virasoro algebras. I
also describe several unanswered problems and some proposals for how
these might be addressed.
\end{abstract}

\pacs{04.70.Dy, 11.25.Hf}

\maketitle

\newpage

\section{Introduction}

Recently, exotic black holes whose masses and angular momenta are
interchanged have been found in several different systems. These are
(a) asymptotically anti-de Sitter black holes in 2+1 dimensional
gravity for the case of a vanishing cosmological constant with
minimally coupled topological matter, which is called `` BCEA ''
gravity \cite{Carl:91}, (b) constant curvature black holes in 4+1
dimensional anti-de Sitter space \cite{Bana:98}, and (c) BTZ-like
black holes in gravitational Chern-Simons theory
\cite{Dese:82,Kalo:93,Cho:98,Garc:03,Krau:05,Solo:05,Saho:06}. But,
it is known that these black holes do not satisfy the
Bekenstein-Hawking entropy formula, but depend only on the
area of the $inner$ horizons, %$r_{-}$
 in order to satisfy the first
law of thermodynamics. This looks similar to Larsen's suggestion in
another context \cite{Lars:97}. But, a basic problem of these
approaches is that the second law of thermodynamics is not
guaranteed with their entropy formulae,
%for the inner horizon area
in contrast to the Bekenstein-Hawking form \cite{Hawk:71}. Actually,
without the guarantee of the second law, there is no justification
for identifying entropies with the inner horizon areas
\cite{Beke:73}.

In the usual system of black holes, the first law of thermodynamics
uniquely determines ( up to an arbitrary constant ) the black hole
entropy with a given Hawking temperature $T_{H}$ and chemical
potential for the event horizon $r_+$. In this context, there is no
choice in the entropy for the exotic black hole, other than
proportional to the area of the inner horizon $r_-$.
%In contrast to this,
In this paper, I show  that there is another rearrangement of the
first law such as the entropy has the usual Bekenstein-Hawking form,
but now the characteristic temperature and chemical potential are
those of the inner horizon, in contrast to the previous approaches.
And the temperature has a $negative$ value, due to an upper bound of
mass as in spin systems, and the angular velocity has a $lower$
bound. It is not yet clear how to measure these characteristics by a
physical observer who is in the outside of the event horizon. But, I
show that one can obtain the same entropy from a conformal field
theory computation, based on classical Virasoro algebras at the
spatial infinity.

\section{Thermodynamics}
\label{2}

The three systems
\cite{Dese:82,Kalo:93,Cho:98,Garc:03,Krau:05,Solo:05,Saho:06} which
I have mentioned in the Introduction look different physically. But,
they all allow the exotic black hole solution with the following
properties.

a. It has the same form of the metric as the BTZ (
Banados-Teitelboim-Zanelli ) solution \cite{Bana:92}, or modulus an
expanding/contracting 2-sphere for the case of `(b)',
\begin{eqnarray}
\label{BTZ}
 ds^2=-N^2 dt^2 +N^{-2} dr^2 +r^2 (d \phi +N^{\phi}
dt)^2
\end{eqnarray}
with
\begin{eqnarray}
N^2=\frac{(r^2-r_+^2) (r^2-r_-^2)}{l^2 r^2},~~ N^{\phi}=-\frac{r_+
r_-}{l r^2}.
\end{eqnarray}
Here, $r_+$ and $r_-$ denotes the outer and inner horizons,
respectively.

b. But, its mass and angular momentum are interchanged as
\begin{eqnarray}
\label{M_J}
 M=x j/l, ~~ J=x l m \label{MJ}
\end{eqnarray}
with an appropriate coefficient $x$: $x=1$ for the BCEA black hole
\cite{Carl:91}, $x$ is a fixed value of $U(1)$ field strength for
the case of `(b)' \cite{Bana:98}, and $x$ is proportional to the
coefficient of the gravitational Chern-Simons term for the case of
`(c)'.
% \cite{Solo:05}.
Here, $m$ and $j$ denote the usual mass and angular momentum for
the BTZ black hole
\begin{eqnarray}
\label{mj_BTZ}
 m=\frac{r_+^2 +r_-^2}{8 Gl^2},~~j=\frac{2 r_+
r_-}{8Gl }
\end{eqnarray}
with a negative cosmological constant $\Lambda=-1/l^2$. One
remarkable result of (\ref{M_J}) is that
\begin{eqnarray}
\label{M_bound}
 (l M)^2 -J^2 =x^2 [j^2 -(lm)^2] \leq 0
\end{eqnarray}
for any non-vanishing $x$, which shows an upper bound for the mass
$M$, with a saturation by the extremal case of $j^2=(lm)^2$.

c. On the other hand, since it has the same form of the metric as
the BTZ solution, it has the same form of the Hawking temperature
and angular velocity of the event horizon $r_+$ as in the BTZ also
\begin{eqnarray}
T_+=\left. \frac{\hbar \kappa}{2 \pi} \right|_{r_+}=\frac{\hbar
(r_+^2 -r_-^2)}{2 \pi l^2 r_+},~~\Omega_+=\left.-N^{\phi}
\right|_{r_+}=\frac{r_-}{l r_+}
\end{eqnarray}
with the surface gravity function $\kappa={\partial N^2}/{2
\partial r}$. Now, by considering the first law of thermodynamics as
\begin{eqnarray}
\delta M=\Omega_+\delta J + T_+ \delta S
\end{eqnarray}
with $T_+$ and $\Omega_+$ as the characteristic temperature and
angular velocity of the system, one can easily determine the black
hole entropy as
\begin{eqnarray}
S=x \frac{2 \pi r_-}{4 G \hbar}.
\end{eqnarray}
There is no other choice in the entropy in this usual context
\cite{Carl:91,Bana:98,Solo:05,Saho:06}. But, a basic problem of this
approach is that the second law of thermodynamics is not guaranteed
with the entropy formula, which depends only on the inner-horizon
area $A_-=2 \pi r_-$: Some of the assumptions for the Hawking's area
theorem, i.e., cosmic censorship conjecture might not be valid for
the inner horizon in general. Moreover, the usual instability of the
inner horizon makes it difficult to apply the Raychaudhuri's
equation to get the area theorem, even without worrying about other
assumptions for the theorem; actually, this seems to be the
situation that really occurs in our exotic black holes also
\cite{Stei:94,Bala:04}.

Now, without the guarantee of the second law of thermodynamics,
there is no justification for identifying entropy with the inner
horizon area, even though its characteristic temperature and angular
velocity have the usual identifications \cite{Beke:73}. So, in order
to avoid this problem, we need another form of the entropy which is
$linearly$ proportional to the outer horizon area $A_+=2 \pi r_+$,
following the Bekenstein's general argument \cite{Beke:73}, which
should be valid in our case also, but then the first law would be
satisfied with some another appropriate temperature and angular
velocity. After some manipulation, one finds that the first law can
be actually rearranged as
\begin{eqnarray}
\label{SL:new}
 \delta M=\Omega_-\delta J + T_- \delta S_{new}
\end{eqnarray}
with the black hole entropy
\begin{eqnarray}
S_{new}=x \frac{2 \pi r_+}{4 G \hbar} \label{BH_new}
\end{eqnarray}
and the characteristic temperature and angular velocity
\begin{eqnarray}
T_-=\left. \frac{\hbar \kappa}{2 \pi} \right|_{r_-}=\frac{\hbar
(r_-^2 -r_+^2)}{2 \pi l^2 r_-},~~\Omega_-=\left.-N^{\phi}
\right|_{r_-}=\frac{r_+}{l r_-}
\end{eqnarray}
for the inner horizon. Here, I note that the entropy (\ref{BH_new}),
for the BCEA gravity \cite{Carl:91}, gives the exactly the same
factor as the usual Bekenstein-Hawking formula, but it depends on
other parameters in general
% and is not necessarily the same as the Bekenstein-Hawking formula
\cite{Bana:98,Solo:05,Saho:06}.

 With this new formulation, we
have a dramatic departure from the usual situations. First, the
angular velocity has a lower bound $\Omega_- \geq 1/l$ due to the
fact of $r_+ \geq r_-$; it is saturated by the extremal case
$r_+=r_-$ and divergent for the vanishing inner horizon. This
implies that this system is always rotating, as far as there is the
event horizon $r_+$. Second, the temperature $T_-$ and the surface
gravity $\kappa_{-}$ have negative values. [ I used the definition
of $\kappa$ as $\nabla ^{\nu} (\chi ^{\mu} \chi_{\mu} )=-\kappa
\chi^{\nu}$ for the horizon Killing vector $\chi^{\mu}$ in order to
determine its sign, as well as its magnitude.] The negative-valued
temperature looks strange in the usual black hole context, but this
is a well-established concept in the spin systems where some $upper~
bound$ of the energy level exists \cite{Kitt:67}. Actually, this is
exactly the same situation as in our case, due to the upper bound of
mass in (\ref{M_bound}), and this provides a physical justification
for introducing the negative temperature in our system
also\footnote{One might consider the positive-valued surface gravity
and temperature with $T=\left| \kappa_{-} /(2 \pi) \right| $ (as in
\cite{Bala:04}), but in this case one has an incorrect sign in front
of the $TdS$ term in (\ref{SL:new}).}. This would be probably the
first example in the black hole systems where the negative
temperature occurs.

\section{Statistical Entropy}

It is well known that the black hole entropy for the BTZ black hole
can be also computed statistically using conformal field theory
results \cite{Stro:98,Bana:99}. So it is natural to expect the
similar things in our case also since one has the same form of the
metric as in the BTZ. Here I consider, in particular, the case of
gravitational Chern-Simons gravity
\cite{Dese:82,Kalo:93,Cho:98,Garc:03,Krau:05,Solo:05,Saho:06} which
has been interested recently in the context of higher curvature
gravities also \cite{Krau:05,Solo:05,Saho:06} and whose conformal
field theory  analysis is evident; however, I suspect the similar
results for the cases of `(a)' and `(b)' \cite{Carl:91,Bana:98}
also, although the explicit realizations would be different. There
are several approaches to compute the statistical entropy from
conformal field theory. Here, let me consider the Chern-Simons gauge
theory approach in this paper.

To do this, I first note the equivalence of
\begin{eqnarray}
I_{CS}[A^+]-I_{CS}[A^-]=I_{GCS} [e, \omega]
\end{eqnarray}
for the Chern-Simons gauge action and the gravitational Chern-Simons
action \cite{Witt:88},
\begin{eqnarray}
&&I_{CS}[A^{\pm}]=\pm \alpha \frac{k}{4 \pi} \int d^3 x \left<
A^{\pm}
(d A^{\pm}+\frac{2}{3} A^{\pm} A^{\pm} ) \right>, \nonumber \\
&&I_{GCS}[e, \omega]=-\frac{\alpha}{32 \pi G} \int d^3 x \left<
\omega (d \omega+\frac{2}{3} \omega \omega ) +\frac{e}{l^2} T
\right>,
\end{eqnarray}
respectively, with $A^{\pm}=A^{\pm}_a J^a =(\omega_a \pm e_a/l
)J^a,~\left< J_a J_b \right> =(1/2) \eta_{ab}~ [
\eta_{ab}=diag(-1,1,1) ]$, and $T=de +2 \omega e$ is the torsion
2-form. Then, it is easy to see that the BTZ solution (\ref{BTZ})
satisfies the equations of motion of gravitational Chern-Simons
action $C^{\mu \nu}=0$ with the Cotton tensor $C^{\mu \nu}=
\epsilon^{\mu \rho \sigma} \nabla_{\rho}
(R^{\nu}_{\sigma}-\delta^{\nu}_{\sigma} R/4 )/\sqrt{g} $
\cite{Dese:82}.

Now then, it is straightforward to apply the usual result of Ref.
\cite{Bana:99}, where the Virasoro algebras with {\it classical }
central charges are obtained, since the whole computation is
governed by the properties of BTZ solution (\ref{BTZ}) only. In this
way, one can obtain ( see Ref. \cite{Park:06} for the details ) two
set of Virasoro algebras for the asymptotic isometry group
$SL(2,{\bf R}) \times SL(2,{\bf R})$ with the classical central
charges
\begin{eqnarray}
c^{\pm}=\gamma^{\pm} \frac{3 l}{2G}
\label{c}
\end{eqnarray}
with $\gamma^{\pm}=\pm {\alpha}/{4 l}$ and the ground state
generators
\begin{eqnarray}
L_0^{\pm}=\gamma^{\pm} \frac{1}{2} (lm \pm j) +\frac{c^{\pm}}{24}.
\label{L0}
\end{eqnarray}
Note that, if one identifies the first term in (\ref{L0}) with
$(lM\pm J)/2$ as in the BTZ  case \cite{Krau:05,Solo:05,Saho:06},
one finds that $M$ and $J$ are identified with those of (\ref{MJ})
with $x=\alpha/(4 l)$; however, my computation based on conformal
field theory does not depend on the manner of identifications of $M$
and $J$, but only on $r_+$ and $r_{-}$. With the data of (\ref{c})
and (\ref{L0}), one can now compute the statistical entropy from the
Cardy's formula for the asymptotic states \cite{Carl:99} as
\footnote{If I consider the system with both the Einstein-Hilbert
term as well as the gravitational Chern-Simons term as in Ref.
\cite{Solo:05}, there is the inner-horizon's contribution also, in
general. My result can be obtained from the general formula by
considering $|\beta|/l \rightarrow \infty$ limit, where the
inner-horizon's contribution is negligible. However, the resulting
formula (5.7) of Ref. \cite{Solo:05} does not do the job,
%even though it was deduced from the Cardy's formula.
and this is basically because it is valid only for $|\beta|/l<1$
\cite{Park:06}.}
\begin{eqnarray}
S_{stat}&=&\frac{2 \pi}{\hbar} \sqrt{\frac{1}{6} \left(c^+
-L_{0(min)}^+\right)\left(L_0^+ -\frac{c^+}{24} \right)} + \frac{2
\pi}{\hbar} \sqrt{\frac{1}{6} \left(c^-
-L_{0(min)}^- \right)\left(L_0^- -\frac{c^-}{24} \right)} \nonumber \\
&=&\frac{2 \pi r_+}{4 G \hbar} \left| \frac{\alpha}{4 l} \right|,
\end{eqnarray}
where I have chosen $L_{0(min)}^{\pm}=0$ for the minimum value of
$L_0^{\pm}$ as usual \cite{Stro:98}; this corresponds to the $AdS_3$
( three-dimensional anti-de Sitter space ) vacuum solution in the
usual context, but it has a permanent rotation with the angular
momentum $J=-({\alpha}/{2 }) ({l}/{16 G})$ and the vanishing mass
$M=0$ in our new context \cite{Krau:05}.

So, one finds an exact agreement for the case of $\alpha >0$, where
$M,~J,$ and $S_{new}$ are positive definite, with my new entropy
formula (\ref{BH_new}). Hence, the new entropy formula for the
exotic black holes is supported by the statistical computation,
based on conformal field theory. Note that, in this case, all
$c^{\pm}$ and $L_0^{\pm}-c^{\pm}/24$ are not positive definite, but
their self-compensations of the negative signs produce the positive
entropy\footnote{The application of the Cardy's formula to the case
of negative $c$ and $L_{0}$ might be questioned due to the existence
of negatives-norm states with the usual condition $\left.
L_n|h\right>=0~(n>0)$ for the highest-weight state
$\left.|h\right>$. However, this problem can be easily cured by
considering another representation of the Virasoro algebra with
$\hat{L}_{n}\equiv -L_{-n},~\hat{c}\equiv-c$, and $\hat{L}_n|\hat{h}
\left.\right>=0~(n>0)$ for the new highest-weight state
$|\hat{h}\left.\right>$ \cite{Bana:99b}. So, the formula (16), which
is invariant under this substitution, can be understood in this more
precise context also.}. But for the case of $\alpha <0$, where
$S_{new}$, as well as $M$ and $J$, becomes negative, the statistical
counterpart does not exist in principle, from its definition
$S_{stat}=ln \rho \geq 0$ for the number of possible states $\rho
(\geq 1)$. So, it is not so surprising that we have found a
disagreement in this latter case.
%One might consider a positive temperature $\tilde{T}_-=|T_-|$ such
%as the entropy becomes positive as
%$\tilde{S}_{new}=\frac{\alpha}{4 l^2} \frac{2 \pi r_+}{4 G
%\hbar}$, which agrees to the statistical entropy, since (negative)
%mass $M$ has a lower bound $J/l$, rather than the upper bound. But
%the problem is that the second law of thermodynamics is still
%questionable due to violation of the (null) energy condition from
%the negative mass.

\section{Summary and Discussion}

I have argued that even the exotic black holes with the interchanged
masses and angular momenta have the black hole entropies with the
usual Bekenstein-Hawking form, but now their characteristic
temperatures and angular momenta are those of the inner horizons. I
have found that the new entropy formula agrees with the statistical
entropy, based on classical Virasoro algebras at the asymptotic
infinity. In the statistical analysis I have considered only the
case of gravitational Chern-Simons gravity, and it is believed that
similar results would be obtained for the other two cases also.
%But their explicit computations would be clearly of interest. And
%also, a direct measurement of $T_-$ and $\Omega_-$ by some
%physical devices would be a challenging problem; the $AdS/CFT$
%probing beyond the event horizon \cite{Bala:04} might be a
%possible device but this needs further studies.
But, there are still several unanswered problems, and I will below
describe the problems and some possible proposals for how these
might be addressed.\\

1. We know that black holes are thermal objects because they emit
Hawking radiation with a thermal spectrum.  In the standard analysis
initiated by Hawking \ci{Hawk:75}, this spectrum is determined by
the metric alone. However, this work implies that two black holes
with identical BTZ metrics will emit radiation with different
spectra, one a black body spectrum corresponding to a positive
temperature $T_+$ for the ordinary black hole and one a very
non-black-body spectrum corresponding to a negative temperature
$T_-$ for the exotic black holes. Then, {`` Can we give a plausible
explanation of why the standard computation of black hole
temperature should fail in the exotic cases ? ''} And { `` How can
we compute the Hawking radiation if the standard computation
fails ? ''}\\

This would be the most important but the most difficult question
whose complete answer is still missing. But here, I would like to
only mention the possible limitation of the standard approach in the
exotic black hole case and how this $might$ be circumvented. To this
end, I first note that, in the standard computation of Hawking, the
background metric is considered fixed such as the back-reaction
effects are neglected. Now, the question is how much we can trust
this approximation to get the {\it leading} Hawking radiation
effects for the real dynamical geometry ? In order to clarify this,
let me consider a black hole with `` rotation ''. Then, I note that
we need to choose an appropriate coordinate, called {\it
co-rotating} coordinate, with the condition $\widetilde{N}^{\phi}
\equiv N^{\phi}+ \Omega_+ \equiv 0$ at the `` outer '' horizon $r_+$
in order to have a well-defined analysis, i.e, analyticity, {\it
near} the outer horizon \ci{Cart:68,Hart:76}, where the Hawking
radiation occurs. And also this makes the $s$-wave or WKB
approximation to be justified \ci{Viss:01} since the radial wave
number approaches infinity near the horizon due to the coincidence
of the infinite redshift surface and outer horizon, even for a
rotating black hole. Now, let me turn to the `` dynamical ''
geometry where the back-reaction effects during the emission process
are considered. Then, it is easy to see that, for the emitted
particles {\it without} carrying the angular momentum, the standard
computation with a fixed background is perfectly well-defined `` at
the initially fixed horizon $r_{+(i)}$ '', though the actual outer
horizon shrinks dynamically at the loss of the emitted positive
energy: With the initial choice of the co-rotating angular velocity
$\Om_+$, one has still $\wt{N}^{\phi}=0$ at the initially fixed
horizon $r_{+(i)}$ such as the infinite redshift surface agrees with
the initial horizon in the co-rotating coordinate system. However,
when there is a change of angular momentum, the situation is quite
different. Actually, in this case there is a {\it finite} separation
of the infinite redshift surface and the initial horizon if we take
into account the loss of the angular momentum, i.e.,
$\wt{N}^{\phi}|_{r_{+(i)}}={s}/{2 r_{+(i)}^2}$, due to angular
momentum $s$ of the emitted particles, with the initially chosen
co-rotating angular velocity $\Om_+$. So, in the standard
computation one does not know whether to use the angular velocity
$\Om_+$ before emission, the angular velocity after emission, or
something in between when consider the co-rotating coordinate
system. This problem looks similar to the situation in the near
extremal black holes when determine a thermal temperature
\ci{Pres:91}, but it would be qualitatively different.

Now, let me explain why this might be relevant to the possible
failure of the standard computation for the exotic black holes.
Here, the important point is that, for the exotic black holes, the
emission of energy $\om$ with an initially chosen co-rotating
coordinate system would reduce the black holes's mass $M$ from the
conservation of energy, but this corresponds to the change of the
angular momentum $j$ of (\ref{mj_BTZ}) in the ordinary BTZ black
hole context, due to the interchange of the roles of the mass and
angular momentum as in (\ref{M_J}). This is in sharp contrast to the
case of ordinary black hole. This seems to be a key point to
understand the peculiar Hawking radiation for the exotic black
holes, and in this argument the conservations of energy and angular
momentum, which are not well enforced in the standard computation,
have a crucial role. So in this respect, the Parikh and Wilczek's
approach \ci{ Pari:00}, which provides a direct derivation of
Hawking radiation as a quantum tunneling by considering the global
conservation law naturally, would be an appropriate framework to
study the problem. But before this, we first need to study the
self-gravitating shells with rotation in Hamiltonian gravity for our
exotic black hole system, as a generalization of Kraus and Wilczek's
\ci{Krau:95}. Currently this
is under study. \\

 2. The Green's function methods for determining the temperature
of a black hole require an equilibrium with matter at the
corresponding temperature \ci{Hart:76}.  This work now implies that
the analysis assumes such an equilibrium with `` some exotic
surrounding matter '' which has a negative temperature, with an
upper bound of energy levels as in spin systems: Otherwise, i.e.,
with the ordinary surrounding matter,  the negative temperature
black hole can not be at equilibrium with positive temperature
surroundings since an object with a negative temperature is hotter
than one with any positive temperature. Then, `` How one could build
a container with walls held at a negative temperature in order that
such an equilibrium can exist -- the Universe might have to be
filled with
such `` exotic matter '' ? ''\\

This would be a physically interesting question which might be
relevant to understand our Universe with a dark side. But I suspect
that the resolution would be rather simple in our case from the fact
that in the anti-de Sitter space the artificial container is not
needed in order to study the canonical ( or grand-canonical )
ensemble \ci{Hawk:83,Hawk:99}. But, in the context without the
explicit container, there is a critical angular velocity
\ci{Hawk:99} at which the action of the black hole or the partition
function of its corresponding conformal field theory  diverges.
However, I note that the critical value is the same as the lower
bound of $\Omega_{-}$ such as we are beyond the critical point with
our angular velocity $\Omega_{-}$. So, from this fact, it seems
clear that the ensemble, if there is, in this strong coupling regime
would be quite different from that of the usual BTZ black hole such
as one can not simply apply the usual result to the strong coupling
case. It seems that we need an independent analysis for this case.
But presumably, the ensemble may be still be defined even in the
strong coupling case, due to the symmetry of the BTZ metric under
the $r_{+}\leftrightarrow r_{-}$ exchange.

Finally, I would like to remark that in the standard Green's
function approach the determination of the equilibrium temperature
from the `` fundamental period '', known as the KMS (
Kugo-Martin-Schwinger ) condition \ci{Kubo:57,Mart:59,Lifs:94}, can
be defined without the implicit assumption of a positive
temperature, though not quite well-known in the gravity community (
see Ref. \ci{Brat:76}, for example ). Physicswise, this should also
be the case since the negative temperature is perfectly well-defined
in the ordinary statistical mechanics of spin systems and its
Green's function formulation similarly will reflect the same
temperature, if there is.

\section*{Acknowledgments}

I would like to thank Jacob Bekenstein, Jin-Ho Cho, Gungwon Kang,
Young-Jai Park, and Ho-Ung Yee for useful correspondences. This work
was supported by the Science Research Center Program of the Korea
Science and Engineering Foundation through the Center for Quantum
Spacetime (CQUeST) of Sogang University with grant number R11 -
2005- 021.

%%%%%%%%%% References %%%%%%%%%%%%%%%%%%%%%%%%%
\newcommand{\J}[4]{#1 {\bf #2} #3 (#4)}
\newcommand{\andJ}[3]{{\bf #1} (#2) #3}
\newcommand{\AP}{Ann. Phys. (N.Y.)}
\newcommand{\MPL}{Mod. Phys. Lett.}
\newcommand{\NP}{Nucl. Phys.}
\newcommand{\PL}{Phys. Lett.}
\newcommand{\PR}{Phys. Rev. D}
\newcommand{\PRL}{Phys. Rev. Lett.}
\newcommand{\PTP}{Prog. Theor. Phys.}
\newcommand{\hep}[1]{ hep-th/{#1}}
\newcommand{\hepp}[1]{ hep-ph/{#1}}
\newcommand{\hepg}[1]{ gr-qc/{#1}}
\newcommand{\bi}{ \bibitem}
%%%%%%%%%%%%%%%%%%%%%%%%%%%%%%%%%%%%%%%%%%%%%%%

\end{document}